\documentclass[a4paper]{spie}  
\usepackage[utf8x]{inputenc} 
\usepackage[]{graphicx}
\usepackage{subfigure}
\usepackage{transparent}
\usepackage{amssymb}
\usepackage{amsmath} 
\usepackage[english]{babel}
\usepackage{multicol}

\newcommand{\e}[1]{\times 10^{#1}} 
\newcommand{\ml}[1]{\mathrm{#1}} 

\title{An experimental testbed for NEAT to demonstrate micro-pixel accuracy.} 

\author{Crouzier A. \supit{a}, Malbet F. \supit{a}, Preis O. \supit{a}, Henault F. \supit{a}, Kern P. \supit{a}, Martin G. \supit{a}, Feautrier P. \supit{a}, Cara C. \supit{b}, Lagage P. \supit{b}, Léger A.\supit{c}, LeDuigou J. M. \supit{d}, Shao M. \supit{e}, Goullioud R. \supit{e}
\skiplinehalf
\supit{a}Institut d'Astrophysique et de Planétologie de Grenoble, 414 Rue de la Piscine, St Martin d'Hères, Grenoble, France; \\
\supit{b}Commisariat à l'Energie Atomique et aux Energies Alternatives, Saclay, centre d'études nucléaires de Saclay, Paris, France; \\
\supit{c}Institut d'Astrophysique Spatiale, Centre universitaire d'Orsay, Paris, France; \\
\supit{d}Centre National d'Etudes Statiales, 2 place Maurice Quentin, Paris, France; \\
\supit{e}Jet Propulsion Laboratory, 4800 Oak Grove Drive, Pasadena, CA, U.S.A. 91109
}

\authorinfo{Further author information: (Send correspondence to A.A.A.)\\A.A.A.: E-mail: aaa@tbk2.edu, Telephone: 1 505 123 1234\\  B.B.A.: E-mail: bba@cmp.com, Telephone: +33 (0)1 98 76 54 32}

\begin{document} 
\maketitle 

\begin{abstract}

NEAT is an astrometric mission proposed to ESA with the objectives of detecting Earth-like exoplanets in the habitable zone of nearby solar-type stars. In NEAT, one fundamental aspect is the capability to measure stellar centroids at the precision of $5\e{-6}$ pixel.

Current state-of-the-art methods for centroid estimation have reached a precision of about $4\e{-5}$ pixel at Nyquist sampling. Simulations showed that a precision of 2 µ-pixels can be reached, if intra and inter pixel quantum efficiency variations are calibrated and corrected for by a metrology system.

The European part of the NEAT consortium is designing and building a testbed in vacuum in order to achieve $5\e{-6}$ pixel precision for the centroid estimation. The goal is to provide a proof of concept for the precision requirement of the NEAT spacecraft. In this paper we give the basic relations and trade-offs that come into play for the design of a centroid testbed and its metrology system. We detail the different conditions necessary to reach the targeted precision, present the characteristics of our current design and describe the present status of the demonstration.

\end{abstract}




\section{INTRODUCTION}\label{sec:INTRODUCTION} 

\subsection{Presentation of the NEAT concept}\label{subsec:Presentation of the NEAT concept}

With the present state of exoplanet detection techniques, none of the rocky planets of the Solar System would be detected and indeed their presence is a very strong constraint on the scenarios of the formation of planetary systems. By measuring the reflex effect of planets on their central host stars, astrometry can lead us to the mass of planets and to their orbit determination. This technique is used frequently and is very successful to determine the masses and the orbits of binary stars. However it is necessary to go to space to reach the precision required to detect all planets down to the telluric regime.
\\

\begin{figure}[t]
\begin{center}
\includegraphics[height = 40mm]{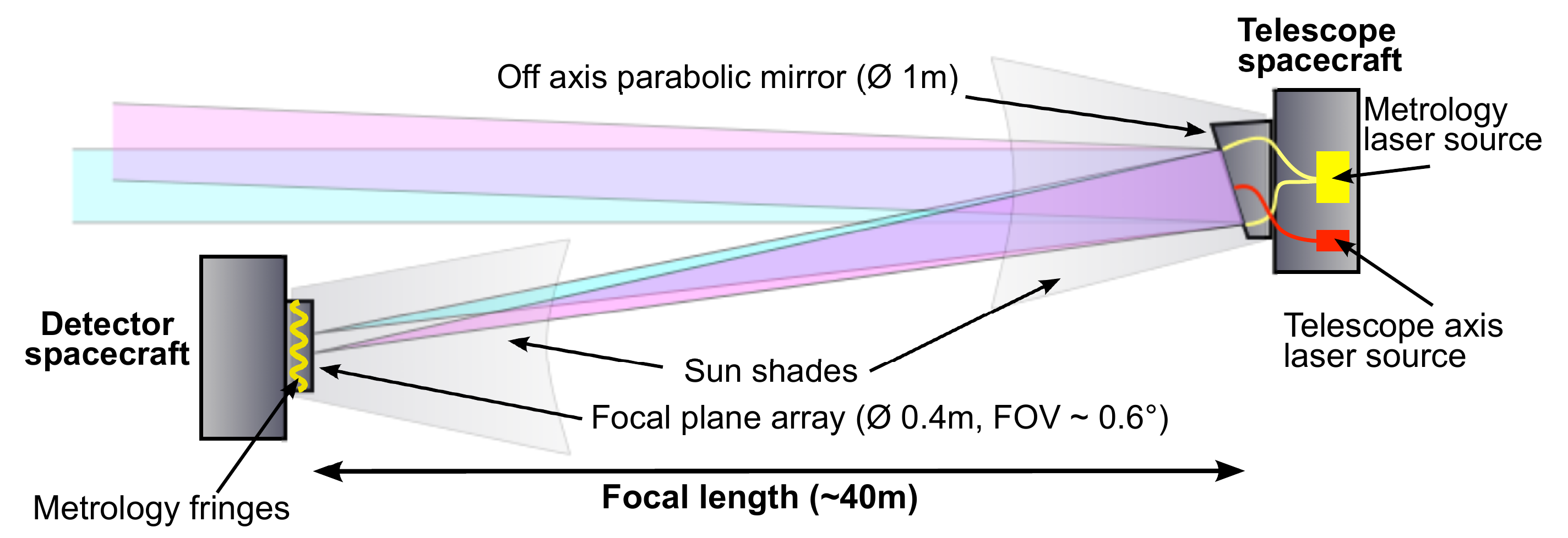}
\caption{\label{neat_concept_diagram}\textbf{The proposed NEAT concept.} The metrology system projects dynamic Young fringes on the detector plane. The fringes allow a very precise calibration of the CCD in order to reach micro-pixel centroiding errors.}
\end{center}
\end{figure}

We have been proposing a mission to ESA in the framework of the call for M missions in the Cosmic Vision plan which objective is to find most of the exoplanets of our Solar neighbourhood\cite{Malbet11,Malbet12}. The objective is to use differential astrometry to complete the measurements obtained by other techniques in order to lower the threshold of detection and characterization down to the level of an Earth mass in the habitable zone of each system. We want to explore in a systematic manner all solar-type stars (FGK spectral type) up to 20 pc from the Sun. The satellite concept is based on formation flying technology with a satellite carrying a single primary mirror and another satellite carrying the focal plane (see Fig.~\ref{neat_concept_diagram}). The measure is done using laser metrology and interferometry.

One of the fundamental aspects of the NEAT mission is the extremely high precision required to detect exo-Earths in habitable zone by astrometry. The amplitude of the astrometric signal that a planet leaves on its host star is given by the following formula:
\begin{equation}\label{eq:astrometric_signal}
A = 3 \mu \ml{as} \times \frac{M_{\ml{Planet}}}{M_{\ml{Earth}}} \times \left(\frac{M_{\ml{Star}}}{M_{\ml{Sun}}}\right)^{-1} \times \frac{R}{1\ml{AU}} \times \left(\frac{D}{1\ml{pc}}\right)^{-1}
\end{equation}

Where $D$ is the distance between the sun and the observed star, $M_{\ml{Planet}}$ is the exoplanet mass, $R$ is the exoplanet semi major axis and $M_{\ml{Star}}$ is the mass of the observed host star. For an Earth in the habitable zone located at 10 pc from the sun, the astrometric signal is 0.3 micro arcseconds (or $1.45\e{-11}$ rad). This is smaller than the precision announced for the Gaia mission (launch scheduled for 2013) which should be 7 $\mu$as, in optimal conditions. With a focal length of 40 meters, and taking into account a required signal to noise ratio\cite{sim_double_blind_test09} of 6 and the required number of measurements per target \cite{neat_number_of_measurements11}, the 0.3 $\mu$as requirement to detect an Earth at 10 pc translates into a need to calibrate the pixelation error to $5\e{-6}$ pixels for each integration, as shown by the NEAT error budget\cite{neat_error_budget11}.
\\

In the following subsections we will present the so-called centroid experiment, which goal is to demonstrate the feasibility of the latter requirement: first we present the theoretical aspects, then we introduce the JPL centroid experiment\cite{Nemati11} and its CNES homologue.

\subsection{Presentation of the centroid experiment: theoretical aspects}\label{sec:centroid xp theoretical aspects}

Before entering further into the subject, we need to define what is measured, how we perform the measurements and how we define the errors we will be dealing with. We measure a distance between the location of several centroids. More precisely, we imitate the differential measurement technique of NEAT by comparing the position of the central target star to a reference frame defined by the position of the surrounding reference stars. The final error of the measure is comparable to the error of the centroid location estimation. A straightforward method to estimate a centroid location is to calculate the barycenter of the intensity of light. But this method is affected by the noise of the pixels located at large distances and prevent a precise estimation. 
\\

That is why the standard approach to precision centroid measurement is to do a least square fit of the PSF: a precise knowledge of the PSF is required. To mitigate this problem, we use a centroid displacement algorithm: one image is used to reconstruct the PSF, then the PSF is resampled at different locations. The location at which the resampled image best matches the second image is the estimation of the displacement between the two images. The key point here is that the PSF can be reconstructed very accurately because it is Nyquist sampled and because the pixels response functions are characterised by the metrology\cite{Nemati11}.
\\

The measure of the location of one centroid is affected by numerous types of errors:

\begin{itemize}
\item \textbf{Photon noise error:} Error due to the finite number of photon constituting the centroid. This is a fundamental limitation and represents the ideal error limit for a given centroid estimation method. A fundamental relation in our experiment is the precision reached for a given number of detected photo-electrons. If we knew the exact position of every photon detected during the integration (as if the pixels were all perfect and infinitely small and the detector had an infinite size), the best estimator of the centroid location would be the average location of all $N_{\ml{ph}}$ photo-electrons, and would have a standard deviation of $\frac{\sigma_0}{\sqrt{N_{\ml{ph}}}}$ (by direct application of the Central Limit Theorem). In this relation $\sigma_0$ is the deviation for one photon, i.e. it is the first moment of the centroid intensity distribution. This has important consequences in the design of our experiment. From this relation, we know that to reach a precision of $5\e{-6}$ pixel, we have to use more than $N_{\ml{ph}} = 7 \e{9}$ photo-electrons. This number will determine the integration time necessary to reach the desired precision, given the flux on the CCD and the technical specifications of the latter. For practical purposes, we plan to be able to integrate in less than one hour: that will impose us strict design constraints.
\item\textbf{Pixelation errors:} Errors related to the pixels. This term designate several types of errors. First, each pixel integrate the signal over a surface, and if the centroid is not properly sampled, it can introduce errors. Moreover, inter pixel variations also introduce some errors, even when the centroid is properly sampled: the pixels have different dark currents, sensitivities, they are not regularly spaced and the quantum efficiency (QE) profile within each pixel is not uniform. While the dark current and the sensitivity variations are usually calibrated with the dark and flat fields, at our level or precision we have to take into account all the parameters.
\item\textbf{Truncation error:} Error caused by the finite size of the integration window: a part of the signal is inevitably lost and this leads to additional errors. 
\end{itemize}

The final \textbf{centroid error}, which is the standard deviation between the estimated and true position of the centroid, depends on all the errors mentioned above and on the centroid estimation algorithm used. If we assume that the errors are uncorrelated and the centroid estimation is optimal, we can express the centroid error as the root sum square of the other errors. This implies that all this sources of errors have to be kept below $5\e{-6}$ pixel. Note that we have not considered position errors relative to the wavefront errors of the optics: as we do not change the relative positions of the mirror and the sources, they will remain identical between the measurements so we will be insensitive to them.
\\

The method used to calibrate the CCD and to use the calibration information to perform the centroid measurements is described in Nemati et al.~(2011)\cite{Nemati11} and Zhai et al.~(2010)\cite{Zhai11}. Detailing the method is out of the scope of this article and we invite the reader to refer to the cited publications for further information.

\subsection{JPL centroid experiment}\label{sec:Presentation of the JPL centroid experiment}

\begin{figure}[t]
\begin{center}
\subfigure[]{\label{fig:JPL_xp_picture}
\includegraphics[width=8.2cm]{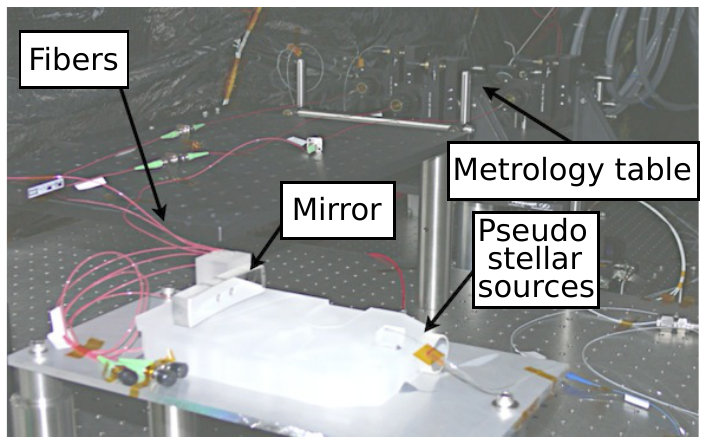}}
\hspace{5pt}
\subfigure[]{\label{fig:JPL_xp_results}
\includegraphics[width=5.8cm]{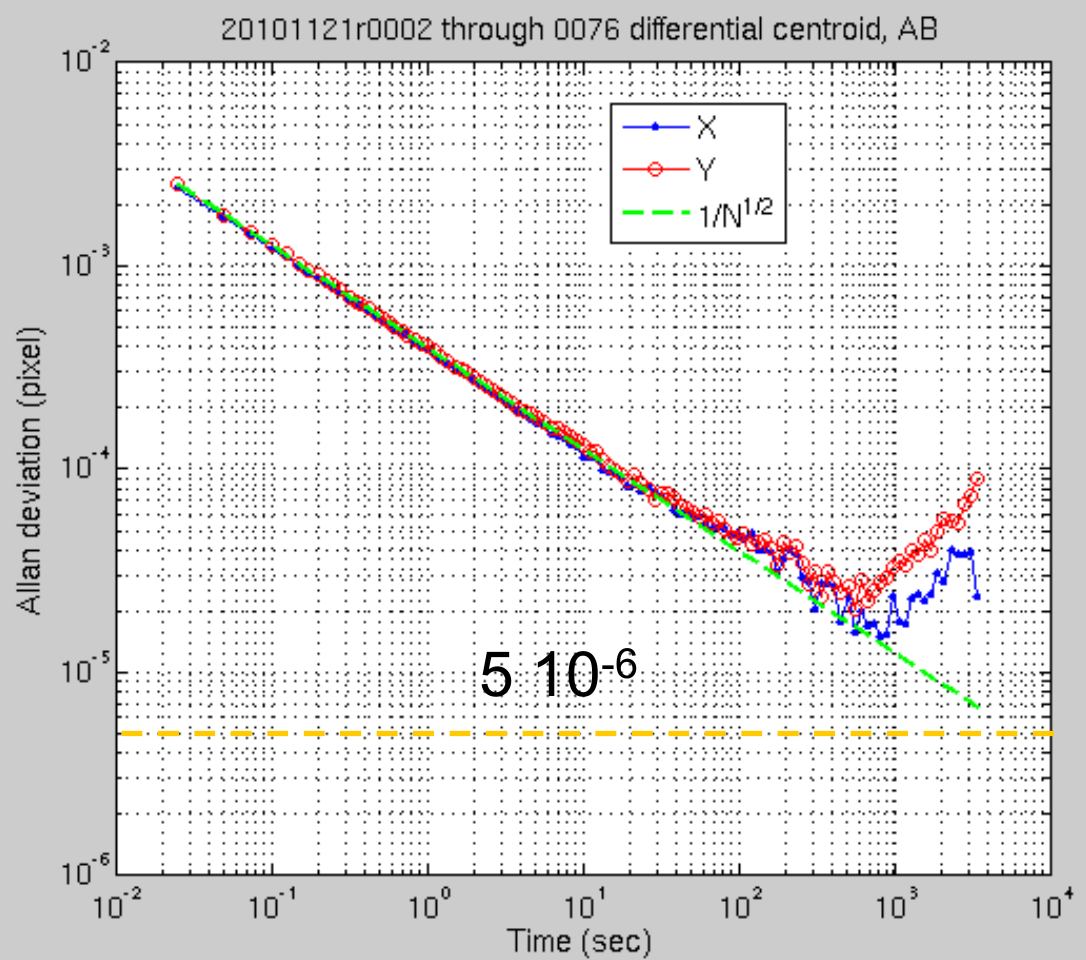}}
\end{center}
\caption{\label{JPL_xp}\textbf{JPL centroid experiment. Left: picture of the experiment. Right: result of the experiment.} The result displayed on the left is the Allan deviation of the measures of the distance between two centroids. One can see that the minimum deviation reached is $4\e{-5}$ pixels for an integration time of a few minutes. Credit: Nemati et al.\cite{Nemati11}}
\end{figure}

An experiment aiming at demonstrating the feasibility of a state-of-the-art calibration at $5\e{-6}$ pixels was conducted at the JPL (and it is still under development). The precision reached was $4\e{−5}$ pixel for Nyquist sampled centroids. Figure~\ref{fig:JPL_xp_picture} is a picture of the testbed and Fig.~\ref{fig:JPL_xp_results} shows the result of the experiment.

The CCD, the sources nor the mirror were moved during the whole integration time, so the dominant source of noise is initially the photon noise, until second order effects becomes dominant, after a few minutes of integration. The increasing values of the Allan deviation are caused by a drift of the centroids on the CCD. This drift is in turn thought to be the result of thermal fluctuations. As the calibration data from the metrology was not used in the centroid-fitting algorithm, a shift of the centroids has introduced errors related to inhomogeneities in pixel characteristics (mainly quantum efficiency and true locations).
\\

During a future phase the CCD will be moved during the integration. By integrating over several sets of pixels the final calibration error should be averaged down. The results of the CCD calibration will be used to refine the fitting method as well.

\subsection{CNES centroid experiment}\label{subsec:Presentation of the CNES centroid experiment}

\subsubsection{Context and timeline}

In order to strengthen the NEAT case in the next ESA call for M class missions, the European part of the NEAT consortium is designing and building a testbed very similar to the one that is used at the JPL. The main difference is that the metrology system will be made of integrated photonic components. The design of the testbed has begun on January 2012 and the components are now being procured. The bench will be assembled in September - October 2012. We expect to get some preliminary results at IPAG before 2013. Tests in vacuum should start in 2013.
\\

The laboratories involved in this project are: the IPAG (Institut d'Astrophysique de Grenoble) - the laboratory where the experiment will initially take place, the CEA (Commisariat à l'Energie Atomique et aux Energies Alternatives) where the electronics for the CCD camera are developed, the IAS (Institut d'Astrophysique Spatiale) where we have the possibility to use a vacuum chamber, the JPL (Jet Propulsion Laboratory) from which the past experience is very valuable to us.
\\

The founding is done by the CNES (Centre National d'Etudes spatiales) and the labex OSUG@2020.

\subsubsection{Presentation of the testbed}

\begin{figure}[t]
\begin{center}
\subfigure[]{\label{fig:centroid_xp_diagram}
\includegraphics[width=7cm]{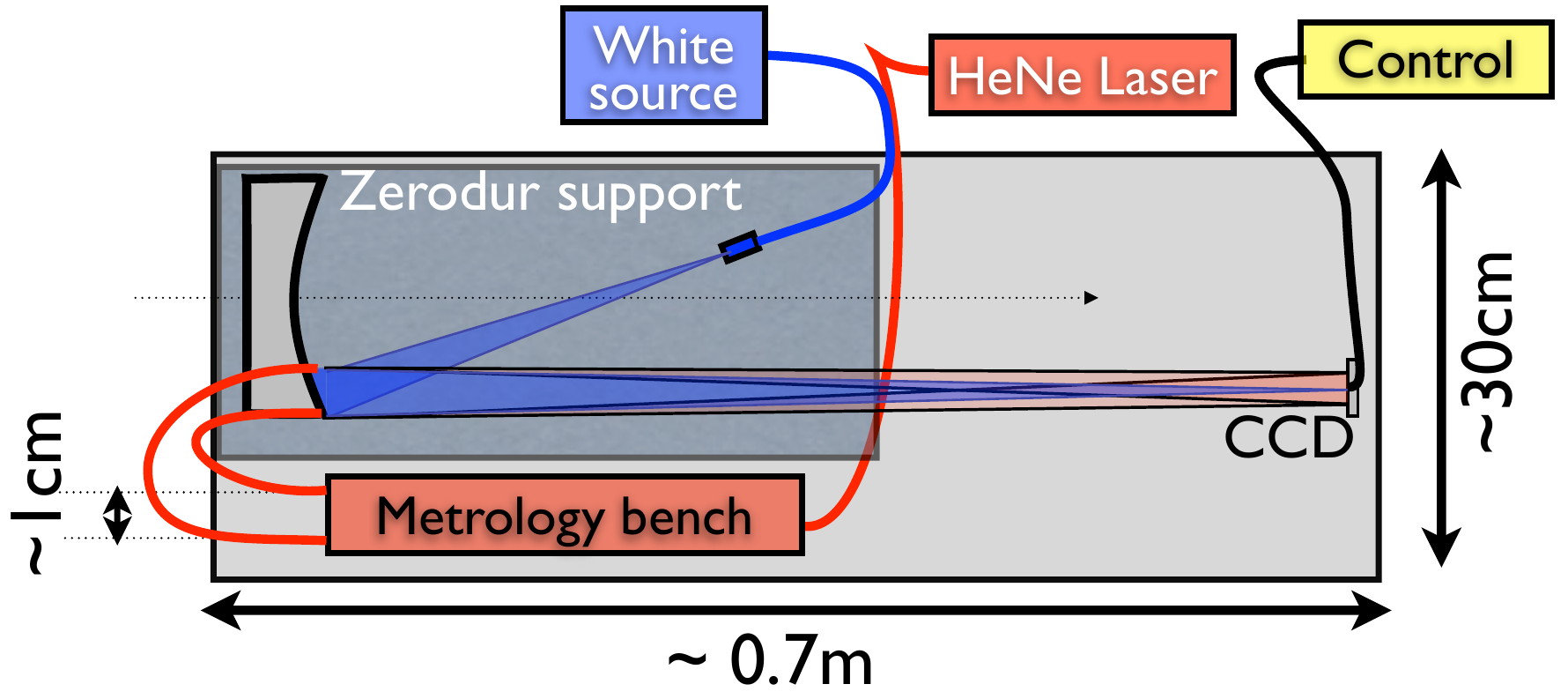}}
\hspace{5pt}
\subfigure[]{\label{fig:centroid_xp_optical_setup}
\includegraphics[width=9cm]{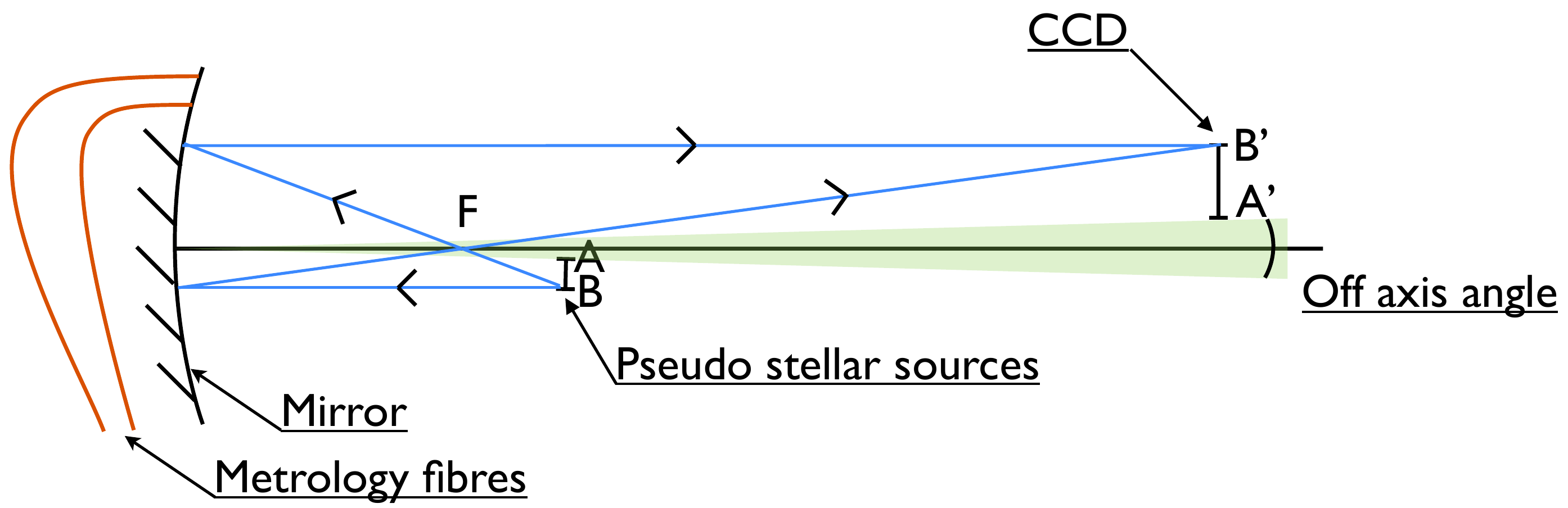}}
\end{center}
\caption{\label{JPL_xp}\textbf{Left: Schematic of the system's components. Right: The optical set-up of the experiment.}}
\end{figure}

The testbed is a simple optical bench that mimics the NEAT optical layout. A spheric mirror images five pinholes which are illuminated by a white source onto a CCD, so that the image is diffraction limited. The five pinholes represent stars, we will refer to them as ‘‘pseudo stellar sources". A set of single-mode fibres located at the edge of the mirror produce laser fringes on the detector. A schematic of the system's components is shown in Fig.~\ref{fig:centroid_xp_diagram}. The optical set-up inside the vacuum chamber is shown in the Fig.~\ref{fig:centroid_xp_optical_setup}.

The most innovative aspect of this experiment is the metrology system that will allow the micro-pixel calibration of the CCD. This system consist of at least two metrology bases (i.e the two pairs of single mode fibres), respectively aligned along the horizontal and vertical axis. The fibre extremities are located next to the mirror and project Young fringes on the detector. Additionally a phase modulator is used to dynamically sweep the fringes over the focal plane. By measuring the intensities variations of the signal for each pixel, one can characterise the inter and intra pixel response of the CCD and bring the centroid error down to the level of a few micro-pixels\cite{Zhai11}.
\\

In the next sections we present the specifications of the NEAT testbed, our design and finally we give the expected performances. For all these sections we will use the notations of the Table~\ref{tab:notations}.

\begin{table}[t]
\hspace{1cm}
\caption{\label{tab:notations}Notations.}
\begin{center}
\begin{tabular}{|l|l|}
  \hline
  Parameter & Notation\\
  \hline
distance mirror to CCD (OA') & $L$\\
minimum/maximum wavelength of the pseudo stellar sources & $\lambda_{\ml{min}} / \lambda_{\ml{max}}$\\
wavelength used for the metrology & $\lambda_m$\\
diameter of the entrance pupil & $D$\\
pixel size & $e$\\
mirror focal length (OF) & $f$\\
separation between the pseudo stellar sources (AB) & $s$\\
metrology baseline & $B$\\
metrology wavelength & $\lambda_{\ml{met}}$\\
  \hline
\end{tabular}
\end{center}
\end{table}

\section{CONSTRAINTS AND SPECIFICATIONS OF THE TESTBED}
\label{sec:CONSTRAINTS AND SPECIFICATIONS OF THE TESTBED}

\subsection{Overview}\label{subsec:Overview}

In order to reach a precision of $5\e{-6}$ pixel, we have to impose ourselves some very strict constraints right from the start. These critical conditions are:

\begin{itemize}\addtolength{\itemsep}{-0.5\baselineskip}
\item Nyquist sampling of the diffraction spot
\item One single optical surface to avoid beam walk errors
\item No beam vignetting
\item Use of a inter and intra pixel calibration system
\item Diffraction limited optics
\item Very high thermal and mechanical stability
\item Low turbulence
\item Large integration zone around the centroids
\end{itemize}

Each one of these conditions is reviewed in the next subsections.


\subsection{Detector constraints}\label{subsec:Detector constraints}

We have to use an off-the-shelve CCD in order to limit costs. The model has to accommodate the following constraints:
\begin{itemize}\addtolength{\itemsep}{-0.5\baselineskip}
\item Nyquist sampling of the PSF (as the size of the mirror will be adjusted to respect this condition, this constraint is in fact relaxed)
\item High ‘‘frame rate $\times$ effective well size", because we want to be able to run the experiment in less than 1 hour.
\item Small dark current at ambient temperature (a few hundreds electrons/pixel/sec) to allow integration over large windows
\item Need to perform series of measures on different CCD positions (i.e. the CCD support has to be motorised and remote controlled)
\end{itemize}

\subsection{Spatial constraints}\label{subsec:Spatial Constraints}

Because we would like to be able to conduct the experiment into a vacuum chamber, we have to put some constraints on the size of the bench. We have the possibility to use the ‘‘Mercure chamber", located at IAS, which is cylindrical and offers a usable space of dimensions:
80 cm (length) x 35 cm (diameter). Some room (at least 10 cm) has to be left to accommodate the CCD electronics into the chamber, leaving a maximum of 70 cm available for the optical bench.
\\
 
\subsection{Optical constraints}\label{subsec:Optical Constraints}

To reach a precision of $5\e{-6}$ pixels, it is necessary to use only one optical surface, otherwise beam walk errors over the multiple optical surfaces will cause systematic errors greater than the targeted precision (we currently do not know how to manufacture multi surfaces optical systems meeting the precision requirement). Using only one optical surface ensure that the systematic error introduced by the wavefront error is identical for all sources located near the optical axis.
\\

The system has to be diffraction limited. If the system is not diffraction limited the constraints will change: first, the PSF will be larger. Furthermore, a micropixel centroid estimation could use up to the first 7 Airy rings, and a geometric diffraction spot could have a different shape. Thus for the proof of concept to be valid it is important to have a diffraction limited system, as it is the case with the space based NEAT design.
\\

The length of the system is constrained by the dimension of the vacuum chamber, so the size of the entrance pupil, which yields the diffraction beam size, is constrained by the size of the pixel. This is because the size of the airy spot is linked to the size of the pupil: $\ml{Spot\;radius} \approx 1.22\frac{L\lambda}{D}$. As a result, to have Nyquist sampling, one must meet the following condition:
\begin{equation}\label{nyquist sampling}
2e < \frac{L\lambda_{\ml{min}}}{D}
\end{equation}
$\lambda_{\ml{min}}$ corresponds to the minimum wavelength for the white source. Because we use white light, we have to have Nyquist sampling at all the wavelengths and the minimum wavelength has to be used. The integral over the photon wavelength is a linear operation, as a result all the wavelengths contribute to the final PSF. Using a “midband” wavelength as reference could result in under-sampling of part of the signal and aliasing. Since the pixel size $e$ is fixed by the choice of the detector, $D$ has to be chosen accordingly.
\\

Additionally the spacing between the pseudo stellar source near the focus must fill some conditions: the spacing cannot be too large (images outside of the detector plane) nor too small (images overlapping). Integrating up to the $7^{th}$ ring translates (at Nyquist sampling) into integrating over 32x32 pixels windows. The minimum separation between the sources depends on the optical magnification $\gamma$, to avoid overlaps between integration zones we must have: 
$$s_{\ml{min}} = \frac{32 e}{\gamma}  = 32e\times\frac{f}{L-f}$$

The coupling of the light from the artificial stars into the system is also an issue. We want to have a uniform intensity distribution on the mirror, meaning the numerical aperture of the pseudo stellar sources has to be larger than the apparent angle of the mirror (seen from the object point). But it should not be much larger than that, or the coupling coefficient will be very low, and the light flux could become an issue.
\\

The geometric aberrations must be small enough, so not to change the PSF significantly. This condition is fulfilled if the off-axis angle is small enough and the optics quality is good enough. The choice of the focal length is important because the it will determine the minimum off-axis angle without obstruction and the aberrations. In practice, since there is no easy analytical way to characterise the aberrations as a function of the focal, we will proceed by trial and error, using ray tracing simulations.

\subsection{Thermal and mechanical constraints}\label{subsec:Thermal and mechanical constraints}

The detector will be chosen to insure a small dark current: we will be able to run the experiment at ambient temperature. Nevertheless, the temperature has to be stabilised in order to insure stability of the bench. The tolerance allowed is very small, because of the extreme precision requirement. The part supporting the mirror and the pseudo stellar sources must be the most stable, it is required that thermal expansion does not change the differential position of the images of the sources of the CCD by more than $5\e{-6}$ pixels. The part supporting the CCD can undergo wider thermal expansions as the position of the CCD is calibrated by the metrology. The mechanical stability requirements are also very tight, for the same reasons. 
\\

The work to quantify the mechanical and thermal stability requirements is still in progress. The current design approach is based on a best effort and on prior experience at the JPL.

\subsection{Metrology constraints}\label{subsec:Metrology Constraints}

The metrology system has to comply with the following list of requirements:
\begin{itemize}\addtolength{\itemsep}{-0.5\baselineskip}
\item Avoid fringing on the detector
\item Nyquist sampling of the Young fringes: $B < \frac{\lambda_{\ml{met}} L}{2e}$
\item Stability of the fringes
\end{itemize}

An important criterion for the choice of the metrology source is the wavelength stability. The fringes must be very stable to allow a precise calibration: the tolerance on the position of the fringe is $5\e{-6}$ pixel. This implies that the laser stability must satisfy:
\begin{equation}
\frac{\delta\lambda_{\ml{met}}}{\lambda_{\ml{met}}}2Ne <  5\e{-6} e \Leftrightarrow \frac{\delta\lambda_{\ml{met}}}{\lambda_{\ml{met}}} < 5\e{-9}
\end{equation} 
Where $N$ is the order of the fringe, here we take $N = 500$: it corresponds to a usable space around the fringe of optical contact of 500 fringes i.e. $500\times 2e = 2.4$ cm.

\section{DESIGN OF THE TESTBED}

\subsection{Bench and optics}

\begin{figure}[p]
\begin{center}
\includegraphics[height = 60mm]{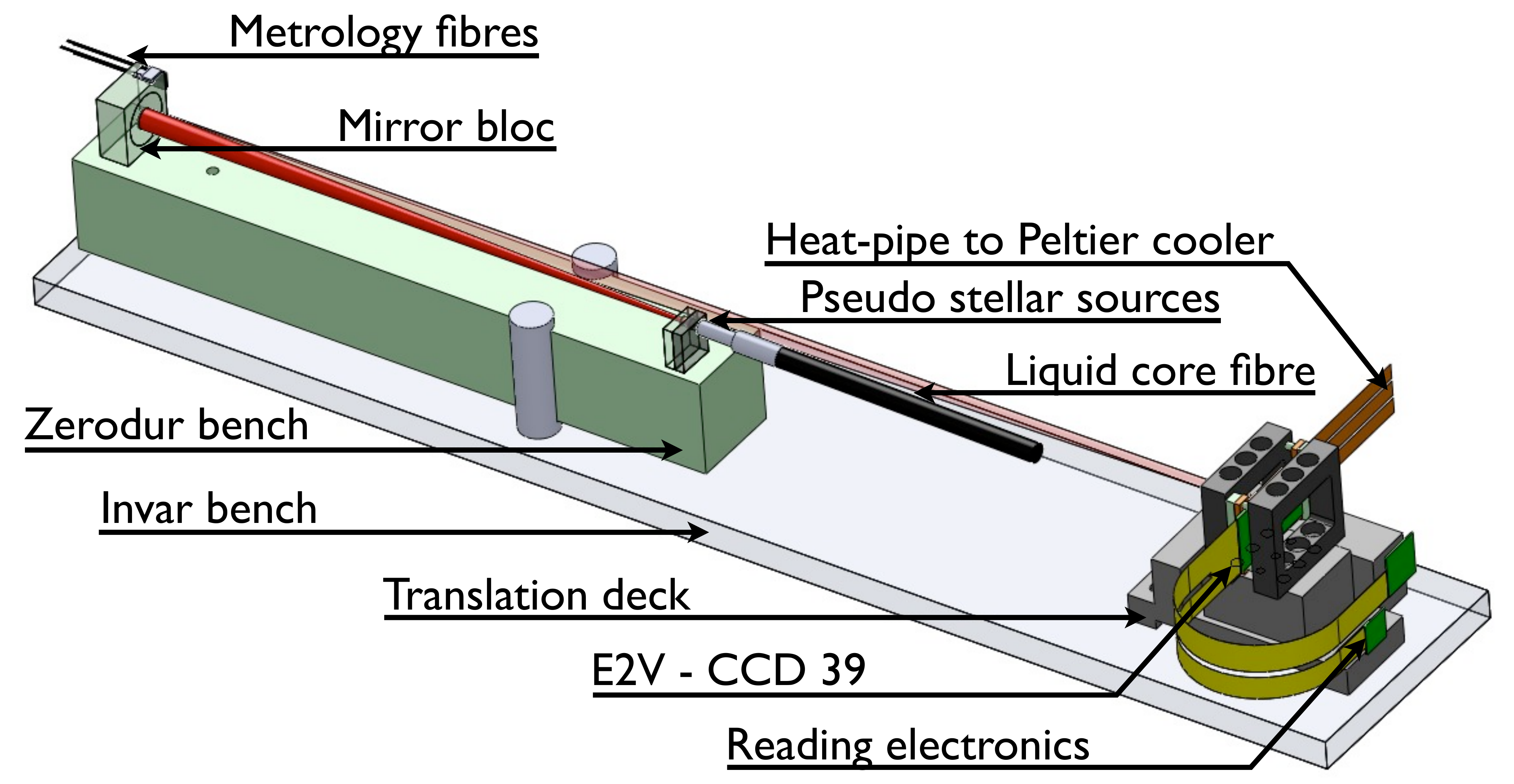}
\caption{\label{fig:centroid_xp_mechanical_concept}\textbf{3D view of the testbed.}}
\end{center}
\end{figure}

Figure~\ref{fig:centroid_xp_mechanical_concept} shows a solidworks 3D view of the testbed. The characteristics of the system we have designed is summed-up in the Table~\ref{tab:optics and mechanical design}.

\begin{table}[p]
\caption{\label{tab:optics and mechanical design}Optical and mechanical designs.}
\begin{center}
\begin{tabular}{|l|l|l|}
  \hline
  Characteristic & Notation & Value\\
  \hline
 Pupil size $= \frac{\lambda L}{2e}$ & $D$ & $5.0$ mm \\
Magnification & $\gamma$ & $2.0$\\
Sources separation & $s$ & $240$ µm\\
Off-axis angle & - & $2\deg$\\
Bench length & $L$ & $0.60$ m\\
Focal length & $f$ &  $0.20$ m\\
Distance OA (pseudo stellar sources to mirror) & $L_s $ & $0.30$ m\\ 
  \hline
\end{tabular}
\end{center}
\end{table}

\begin{figure}[p]
\begin{center}
\includegraphics[height = 80mm]{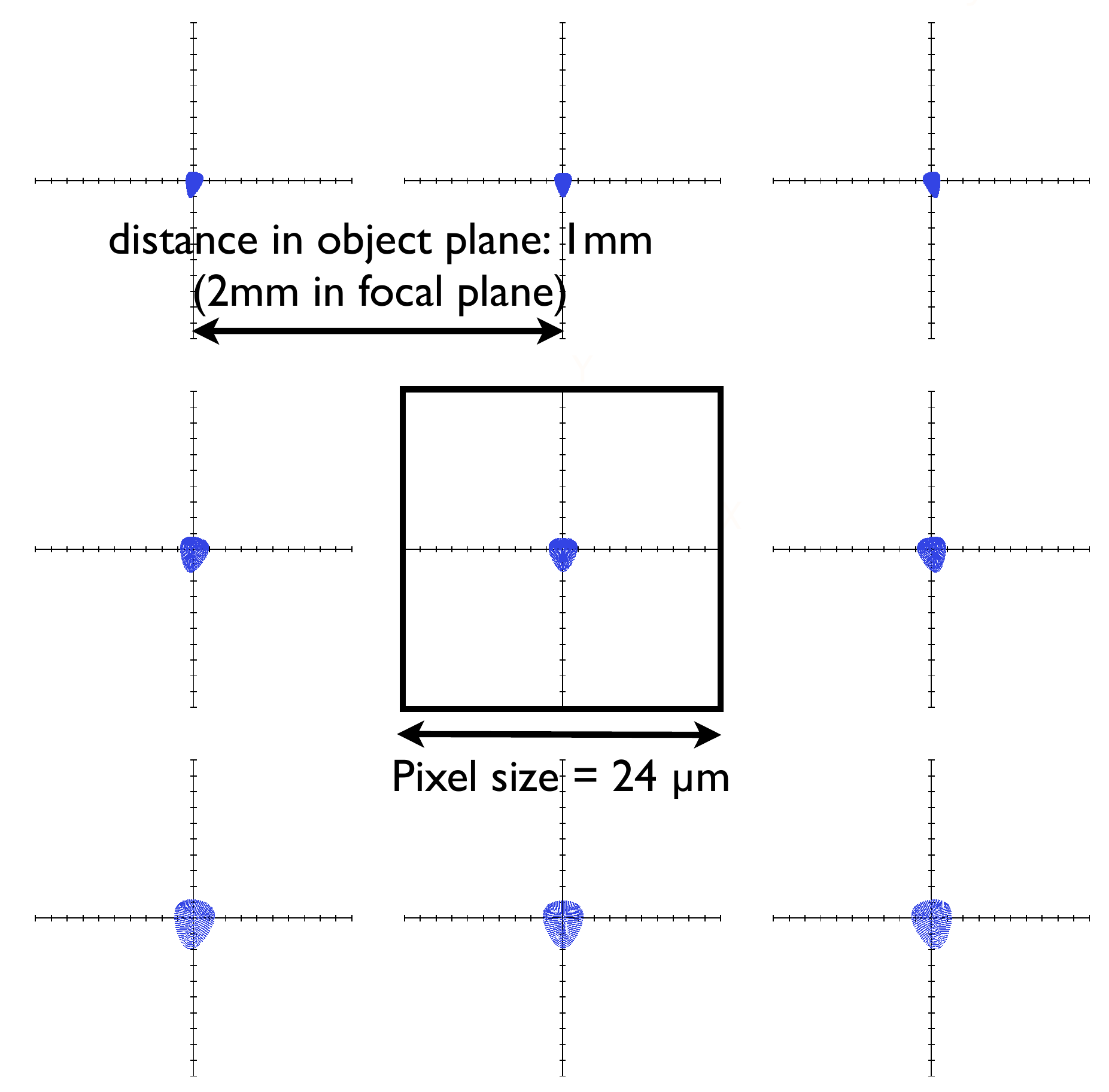}
\caption{\label{fig:spot_diagram}\textbf{Spot diagrams for the chosen optical set-up.} The spot diagrams are spaced by 1 mm in the object plane (it corresponds to a FoV of 0.4 degrees). In our design the sources are only spaced by 240µm, this shows that the aberrations are negligible, within a comfortable margin: in all the cases the size of the spot diagram is a small fraction of the size of the PSF, which has a diameter about 5 pixels.}
\end{center}
\end{figure}

For implementation we used a magnification factor of 2 and an off axis angle of 2 degrees. This configuration allows the installation of the pseudo stellar sources and the camera without any beam obstruction with some margins to accommodate the support elements. Additionally, with an aperture as small as 5 mm, a spherical surface is sufficient to obtain optical aberrations that produce a spot diagram smaller than the diffraction pattern whatever its position in the field of view. The resulting spot diagrams are shown in Fig.~\ref{fig:spot_diagram}. The light source for the pseudo stellar sources is a black body with a temperature of 3000 K.

The current design approach is based on prior experience at the JPL and on a best effort approach, within the allowed budget. The most critical part, i.e. the one that supports the CCD and the pseudo stellar sources (see Fig.\ref{fig:centroid_xp_mechanical_concept})  will be made entirely out of zerodur. Additionally the bench will be thermally regulated to about 0.1 degree. The non critical part (see Fig.~\ref{fig:centroid_xp_mechanical_concept}) supporting the CCD can accommodate larger thermal expansion and will be made in Invar.
\\

A standard suspension table with passive pneumatic supports and an aluminium honeycomb structure will be used. We keep the possibility to add silent blocks between the Invar bench and the table if the vibrations are affecting the measures.

\subsection{Metrology}\label{subsec:Metrology}

\begin{figure}[t]
\begin{center}
\includegraphics[width = 15cm]{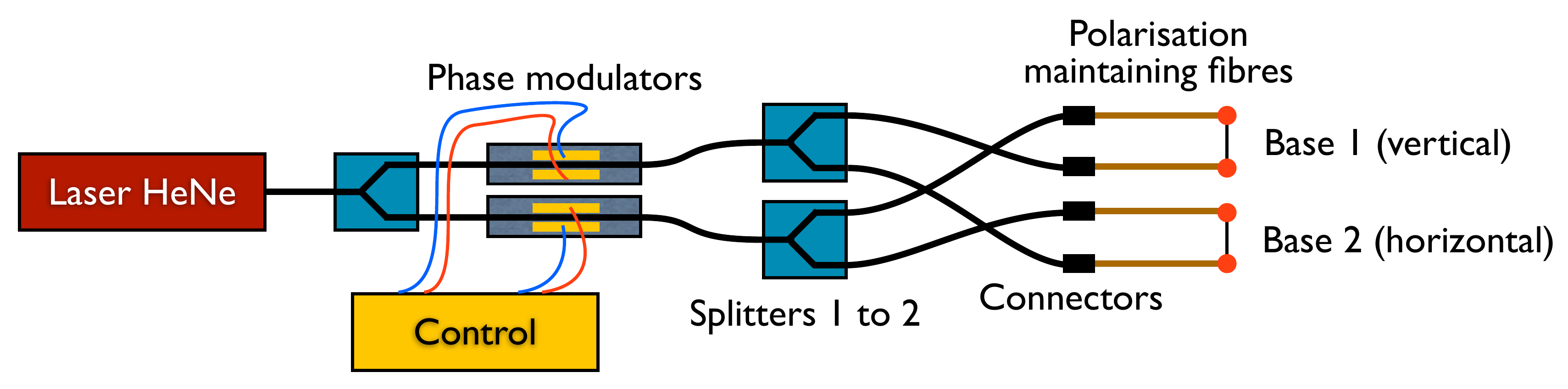}
\caption{\label{fig:metrology_pipeline}\textbf{Schematic of the metrology.} 
}
\end{center}
\end{figure}

The metrology, which is made of integrated components, from the laser to the bases, is shown by Fig.~\ref{fig:metrology_pipeline}. The source for the metrology is a stabilised HeNe laser with a power output of 1.5mW. The frequency stability of the laser is 2 MHz (relative stability of $\delta \lambda / \lambda = 4\e{-9}$, which is consistent with the requirement presented in subsection \ref{subsec:Metrology Constraints}. The light from the laser is fed into the lithium niobate modulators to apply a periodic phase shift between the two lanes. This configuration ensures that the phase modulation is applied between the two sources constituting each base. The shutters are controlled to alternatively block all the metrology sources, during the pseudo stellar source integration phase, or to project either vertical or horizontal dynamic Young fringes during the CCD calibration phase.
\\

The interference pattern created by the horizontal metrology baseline, assuming that the sources are of equal intensity and that the intensity created at the focal plane is uniform is:
$$I(x,y) = 2I_0\left[1+\cos\left(\phi_0 + \Delta \phi (t) + \frac{2\pi xB}{\lambda_{\ml{met}} L}\right)\right]$$
Where $I_0$ is the intensity create at the focal plane by one metrology fibre, $B$ is the metrology baseline, $L$ is the distance between the fibres and the CCD, $\phi_0$ is a static phase difference, $\Delta \phi (t)$ is the modulation applied between the lines and x is the horizontal spatial coordinate (i.e the one aligned with the metrology baseline). Although the exact shape of the fringes is hyperbolic, at the first order the fringes are straight and aligned with the direction perpendicular to the metrology baseline.

\subsection{CCD and electronics}\label{subsec:CCD and electronics}

We chose to use the CCD 39-01 from e2v to take advantage of its high frame rate $\times$ quantum well size, and despite its small matrix size. The characteristics of the CCD (and its electronics) are summarized in Table~\ref{tab:CCD characteristics}.

\begin{table}[t]
\hspace{1cm}
\caption{\label{tab:CCD characteristics}CCD and electronics characteristics.}
\begin{center}
\begin{tabular}{|l|l|}
  \hline
  Characteristic & Value\\
  \hline
Pixel size & $24$ µm\\
Matrix size & $80$x$80$ pixels\\
Read noise & 9 electrons\\
Dark current & 5000 electrons/s @300K\\ 
Effective well size & $200 000$ electrons\\
Read time & 1 ms ($=1$kHz frame rate)\\
Sensitivity range & 0.3 to 0.9 µm\\
  \hline
\end{tabular}
\end{center}
\end{table}

Because the CCD is read at 1kHz, the Poisson noise from the dark current at ambient temperature is only 5 electrons / pixel, which is smaller than the readout noise. This CCD has two buffer zones (masked pixels) that allow a rapid transfer and high frame rate. The high quantum well size and frame rate allow a very fast integration which is highly desirable given the level of photon noise targeted. The performances expected with this detector are detailed in section \ref{sec:EXPECTED PERFORMANCES}.

\section{EXPECTED PERFORMANCES}\label{sec:EXPECTED PERFORMANCES}

\begin{figure}[t]
\begin{center}
\subfigure[]{\label{fig:photometry_white_source}
\includegraphics[width=8cm]{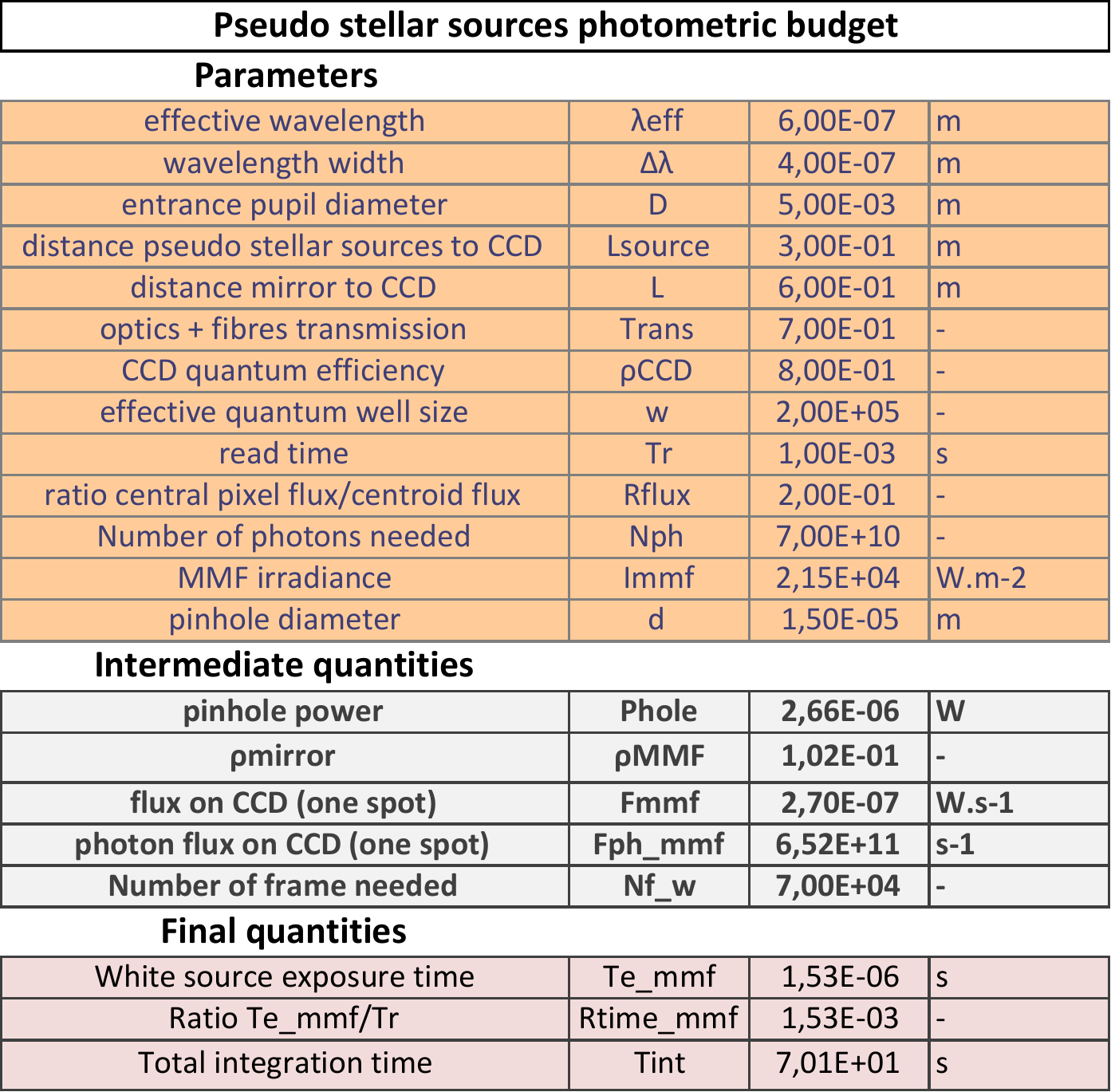}}
\hspace{5pt}
\subfigure[]{\label{fig:photometry_metrology}
\includegraphics[width=8cm]{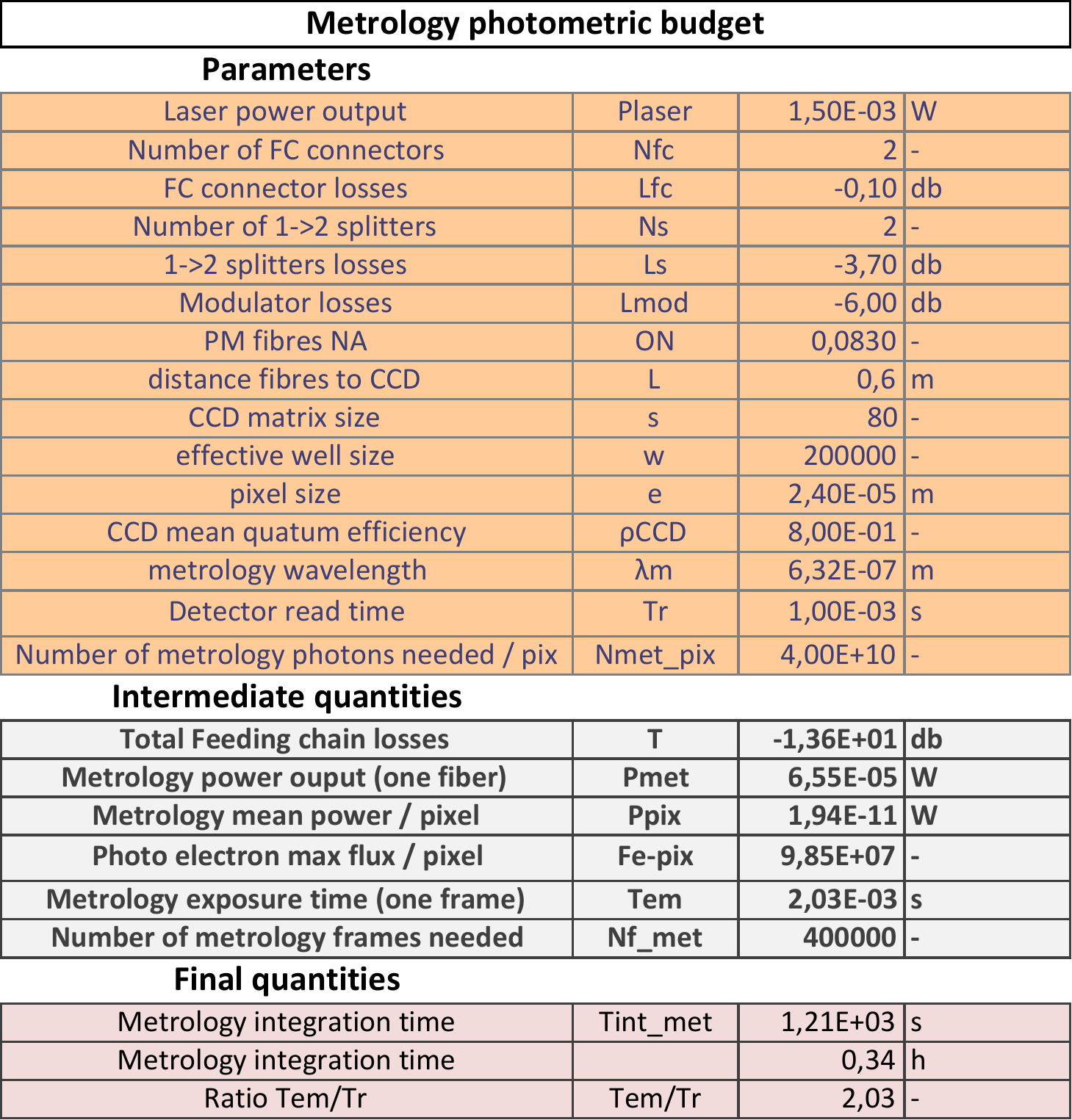}}
\end{center}
\caption{\label{JPL_xp}\textbf{Left: Photometric budget for the pseudo stellar sources. Right: Photometric budget for the metrology.}}
\end{figure}

\subsection{Integration time for the pseudo stellar sources}\label{subsec:Integration time for the pseudo stellar sources}

For this part we assume that we need a number of photons $N_{\ml{ph}} = 10 \times \frac{\sigma_0}{\sqrt{N_{\ml{ph}}}} = 10 \times 7 \e{9} $. Because we want to demonstrate unambiguously that the calibrations errors are smaller than $5\e{-6}$ pixels, we target a centroid error of $5\e{-6}$, so we have to have a photon noise smaller than this value, hence the factor 10.
\\

The total integration time is given by $T_{tot} = N_{\ml{frames}} (T_r + T_e)$, where $T_r$ is the CCD read time and $T_e$ is the exposure time  (i.e. the integration time for one frame). An important criterion to look for is the exposure time: if the flux is large enough, we have an exposure time smaller than the read time of the CCD. In this case the number of photons is sufficient to minimise the integration time, because we have $T_{tot} = N_{\ml{frames}} T_r$ and so the speed is limited by the CCD.

$N_{\ml{frames}}$ is the number of frames needed to reach the number of photons in the centroids, it depends on $N_{\ml{ph}}$, on the size of the quantum wells and on the distribution of the photons in the pixels of the centroid. Our photometric budget showed that, with our set-up and with a black body at 3000 K as the white source, $T_e < T_r$ and the total integration time is 70 seconds.

\subsection{Integration time for the metrology}\label{subsec:Integration time for the metrology}

The error budget is still under development, so the number of photons needed for the metrology measurement, i.e. the characterisation of the pixels intra quantum efficiency profiles, is not well determined yet. But as a rule of thumb, one can assume that the number of photons needed for one pixel is not greater than the one for which the true location of the pixel is determined with a precision better than $5\e{-6}$ pixels. In the centroid fitting algorithm these errors would average down and produce and error smaller than the targeted precision, so this estimate is conservative.
\\

By applying the same formula for the photon noise on one pixel as for the photon noise of the centroid, one find that a number of photons $N_{\ml{ph}} = \left(\frac{e}{5\e{-6}\times e}\right)^2 = 4 \e{10} $ would be sufficient. Our photometric budget showed that the photon flux is critical for the metrology ($T_e > T_r$) and that the integration time is about 20 minutes.
\\

Although the integration time to characterise the CCD is consequent, the integration time necessary to measure the position of the CCD is much lower, because the photon noise in this case is for the CCD as a whole instead of one pixel (6400 times more photons collected). So for several consecutive experiments, assuming constant pixel responses would allow to reduce the metrology integration time dramatically.

\section{Conclusion}

We are in the process of building a testbed that will demonstrate the feasibility of measuring centroids to a precision of $5\e{-6}$ pixel. This will strengthen the case for NEAT as it will show that astrometry down to sub-microarcsec precision is a valid technique for searching Earth-like exoplanets in the habitable zone of nearby stars.

The key specifications of a testbed capable of performing state-of-the-art centroid measurements are: Nyquist sampling of the airy spot, extreme stability of the optics and the use of an interferometric metrology system to calibrate the intra-pixel and inter-pixel quantum efficiency variations.

However, there are still a lot of developments to be made, the most critical ones being the theoretical developments related to the CCD calibration and the centroid estimation algorithms. We are currently developing an error budget that includes mechanical and thermal perturbations, the wavelength, the phase stability and the photon noise of the metrology.

We have presented here a optical bench that has been designed to demonstrate the capability to measure the distance between two sources at the level of 5e-6 pixels. 




\acknowledgments     

~~~~We thank all the engineering team at IPAG that are involved in the project.


We thank the labex OSUG@2020 and CNES for financing the experiment.

At last, we give special thanks to CNES and to Thales Alenia Space for supporting the PhD of Crouzier A.


\bibliography{spie_article_antoine_crouzier}   
\bibliographystyle{spiebib}   

\end{document}